\documentclass{aa}
\usepackage{graphics}

\begin{document}

\title{Hemispheric Sunspot Numbers $R_{n}$ and $R_{s}$:\\
Catalogue\thanks{The catalogue is available in electronic form at
the CDS via anonymous ftp to {\tt cdsarc.u-strasbg.fr} (130.79.128.5)
or via {\tt http://cdsweb.u-strasbg.fr/cgi-bin/qcat?J/A+A/390/707}}
and N-S asymmetry analysis}
\titlerunning{Hemispheric Sunspot Numbers}
\author{M.~Temmer
 \and A.~Veronig
 \and A.~Hanslmeier}

\offprints{M. Temmer, \\ e-mail: {\tt manuela.temmer@uni-graz.at}
}

\institute{Institut f\"ur Geophysik, Astrophysik und
 Meteorologie, Universit\"at Graz, Universit\"atsplatz 5, A-8010
 Graz, Austria}

\date{Received 18 December 2001 / Accepted 6 May 2002}

\abstract{Sunspot drawings are provided on a regular basis at the
Kanzelh\"ohe Solar Observatory, Austria, and the derived relative
sunspot numbers are reported to the Sunspot Index Data Center in
Brussels. From the daily sunspot drawings, we derived the
northern, $R_{n}$, and southern, $R_{s}$, relative sunspot numbers
for the time span 1975--2000. In order to accord with the
International Sunspot Numbers $R_{i}$, the $R_{n}$ and $R_{s}$
have been normalized to the $R_{i}$, which ensures that the
relation $R_{n}+R_{s}=R_{i}$ is fulfilled. For validation, the
derived $R_{n}$ and $R_{s}$ are compared to the international
northern and southern relative sunspot numbers, which are
available from 1992. The regression analysis performed for the
period 1992--2000 reveals good agreement with the International
hemispheric Sunspot Numbers. The monthly mean and the smoothed
monthly mean hemispheric Sunspot Numbers are compiled into a
catalogue. Based on the derived hemispheric Sunspot Numbers, we
study the significance of N-S asymmetries and the rotational
behavior separately for both hemispheres. We obtain that
$\sim$60\% of the monthly N-S asymmetries are significant at a
95\% level, whereas the relative contributions of the northern and
southern hemisphere are different for different cycles. From the
analysis of power spectra and autocorrelation functions, we derive
a rigid rotation with $\sim$27 days for the northern hemisphere,
which can be followed for up to 15~periods. Contrary to that, the
southern hemisphere reveals a dominant period of $\sim$28~days,
whereas the autocorrelation is strongly attenuated after
3~periods. These findings suggest that the activity of the
northern hemisphere is dominated by an active zone, whereas the
southern activity is mainly dominated by individual long-lived
sunspot groups.
}

\maketitle

\section{Introduction}

The relative sunspot numbers $R$ are a measure of solar activity
on the entire disk of the Sun. The relevance of the relative
sunspot numbers lies in particular in the fact that they represent
one of the longest time series of solar activity indices
available. Thus, relative sunspot numbers provide the foundation
of a continuous data set for research on the solar cycle and its
long-term variations. $R$ is defined by
\begin{equation}\label{Wolf}
 R = k \, (10 g + f),
\end{equation}
where $g$ is the number of observed sunspot groups, $f$ the number
of spots and $k$ is an observatory-related correction factor (the
details depending on the actual seeing conditions and the
instrument used).

Historically it can be noted that $R$ is the modified form of the
so-called Wolf index or Wolf number, which was defined without the
correction factor~$k$. The Wolf number was introduced in 1848 by
Rudolph Wolf, who was the first to compile daily Sunspot Numbers.
The original intention of introducing the correction factor $k$ in
about 1882 by Wolf's successors at the Z\"urich Observatory was to
convert the actual daily measurements to the scale originated by
Wolf (cf. Waldmeier~\cite{Wald2}). Waldmeier (\cite{Wald2})
compiled actual records from the Z\"urich Observatory in
coordination with various additional observing stations as well as
data collected by Wolf (\cite{Wolf}). Beginning from 1700, yearly
means of relative sunspot numbers are listed; starting with 1749
monthly mean Sunspot Numbers also are given. This compilation of
the so-called Z\"urich relative sunspot numbers is one of the most
commonly used databases of solar activity (see also
Hoyt~\&~Schatten~\cite{Hoyt1}a,~b). Beginning with 1981, the
Z\"urich relative sunspot program was replaced by the Sunspot
Index Data Center (SIDC) in Brussels, which is the World Data
Center for Sunspot Indices (see also Rishbeth~\cite{Rishbeth};
Ruttenberg~\&~Rishbeth~\cite{Ruttenberg}).

In contrast to the relative sunspot numbers, the hemispheric
Sunspot Numbers $R_{n}$ and $R_{s}$ were not compiled on a regular
basis and are not available before 1992. From January 1992, the
daily $R_{n}$ and $R_{s}$ as well as monthly means have been
compiled by the SIDC (Cugnon~\cite{Cugnon}). However, for several
research purposes, in particular the study of North-South (N-S)
asymmetries of solar activity, as it is presented in this paper
(see Sect.~4), hemispheric Sunspot Numbers are needed for longer
periods. With the present catalogue we aim to provide hemispheric
Sunspot Numbers for at least two full solar cycles (21 and 22). In
general, we derived the hemispheric Sunspot Numbers for the period
1975--2000. The data beginning from January 1992 are used for an
estimation of the consistency of the derived $R_n$ and $R_s$ with
respect to the International $R_n$ and $R_s$ compiled by the SIDC.

At the Kanzelh\"ohe Solar Observatory (KSO), Austria, daily
sunspot drawings have been provided since 1947 within the
framework of solar surveillance programs. The daily provisional
relative sunspot numbers that are derived from the drawings are
compiled and reported to the SIDC in Brussels. The KSO is part of
collaborating observatories all over the world from which
provisional Sunspot Numbers are collected and averaged in an
advanced form. The finally derived numbers are published as the
definitive International Sunspot Numbers $R_{i}$ by the SIDC. A
detailed description of the calculation of the definitive
International Sunspot Numbers, which in particular has to ensure
that the scaling with respect to the Z\"urich Sunspot Numbers is
maintained, is given in Cugnon (\cite{Cugnon}) as well as online
at {\tt http://sidc.oma.be/index.php3}. Details concerning the KSO
and its observing capabilities can be found in Steinegger et al.
(\cite{Steinegger}). Furthermore, the sunspot drawings and other
observational data are online available at the KSO web page at
{\tt http://www.kso.ac.at/}.

The plan of the paper is as follows. In Sect.~2 the gathering and
validation of the hemispheric Sunspot Numbers is described.
Section~3 gives a description of the online catalogue. In
Sect.~4´, an analysis of the N-S asymmetry based on the
hemispheric Sunspot Numbers is presented. In Sect.~5 we discuss
our results and in Sect.~6 the conclusions are drawn.

\section {Data gathering and data validation}

We evaluated the sunspot drawings of the KSO from January 1975 to
December 2000. Within this period, 6900 observation days were
available, which represents a data coverage of $\sim$73\%.
(Missing sunspot drawings are due to bad seeing conditions at the
location of the KSO.) From this data set we derived daily, monthly
mean and smoothed monthly mean hemispheric Sunspot Numbers.
\begin{figure}
\resizebox{\hsize}{!}{\includegraphics{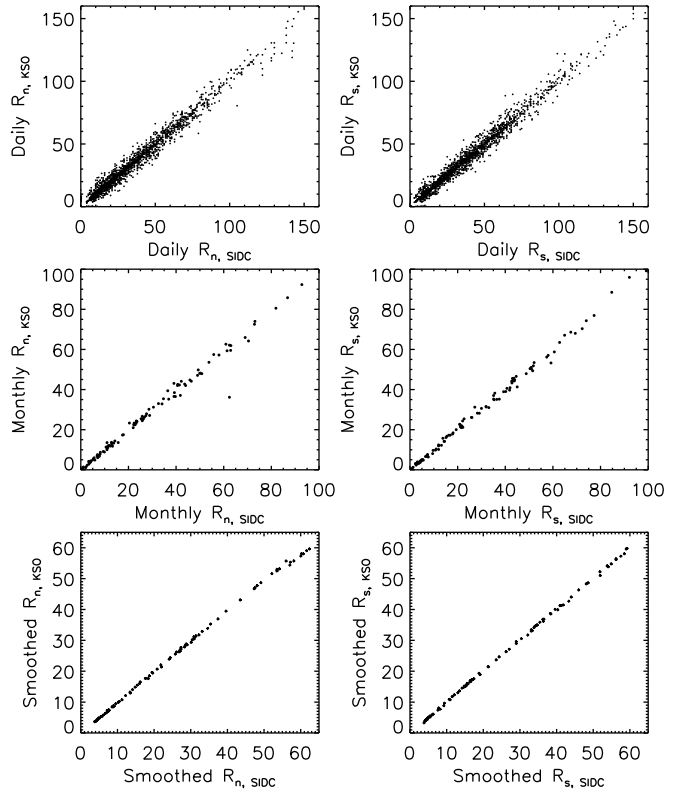}}
    \caption{Scatter plots of the daily (top panels), monthly mean
    (middle panels) and smoothed monthly mean (bottom panels) hemispheric Sunspot
    Numbers for the northern (left panels) and the southern (right
    panels) hemisphere. Calculated hemispheric Sunspot Numbers from KSO, $R_{n {\rm , KSO}}$
    and $R_{s {\rm , KSO}}$, are plotted against the International
    hemispheric Sunspot Numbers provided by the SIDC, $R_{n {\rm , SIDC}}$ and
    $R_{s {\rm , SIDC}}$, for the period 1992--2000.}
    \label{Scatter}
\end{figure}
For each observation day, the Sunspot Number was counted
separately for the northern and the southern hemisphere,
respectively. From this ``raw" hemispheric Sunspot Numbers we
determined the relative fraction of the northern and southern
component, $n$ and $s$. The final hemispheric Sunspot Number,
$R_{n {\rm , KSO}}$ and $R_{s {\rm , KSO}}$, was obtained by
multiplying the northern and southern fractions with the
definitive International Sunspot Number, $R_i$, of the day. With
this procedure we ensure that the derived hemispheric Sunspot
Numbers are normalized with respect to the International Sunspot
Numbers, fulfilling the relation
\begin{equation}\label{norm}
n \times R_{i} + s \times R_{i} = R_{n {\rm , KSO}} + R_{s{\rm , KSO}} = R_{i} \, .
\end{equation}
\begin{figure}
\vspace*{0.15cm}
\center{
 \resizebox{0.9\hsize}{8.5cm}{\includegraphics{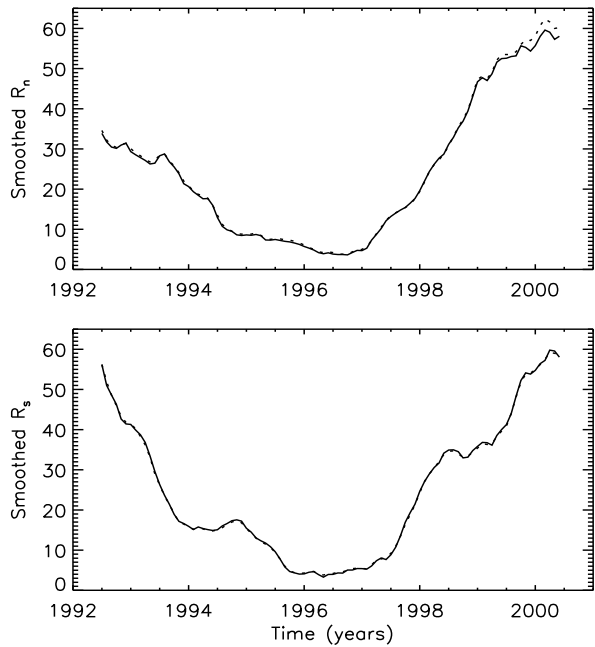}}}
     \caption{Smoothed monthly mean Sunspot Numbers derived from
    KSO data (solid line) and reported from SIDC (dashed line),
    for the northern (top panel) and the southern (bottom panel) hemisphere.}
    \label{Match/365}
\end{figure}
To validate this procedure, we compared the derived hemispheric
Sunspot Numbers, $R_{n {\rm , KSO}}$ and $R_{s {\rm , KSO}}$, with
the International hemispheric Sunspot Numbers, $R_{n {\rm ,
SIDC}}$ and $R_{s {\rm , SIDC}}$, for the overlapping period
1992--2000. For comparison, we utilized the daily, the monthly
mean and the smoothed monthly mean values of the International
hemispheric Sunspot Numbers provided by the SIDC, online available
at {\tt http://sidc.oma.be/}. A detailed description of the SIDC
data sets and the methods that are used for its determination is
given in Cugnon~(\cite{Cugnon}).

The top panels in Fig.~\ref{Scatter} show the scatter plots of the
derived daily Sunspot Numbers, $R_{n {\rm , KSO}}$ and $R_{s {\rm
, KSO}}$, versus the corresponding International Sunspot Numbers
provided by the SIDC, $R_{n {\rm , SIDC}}$ and $R_{s {\rm ,
SIDC}}$, for the period 1992--2000. The scatter plots clearly
reveal that no systematic difference exists between the derived
and the International daily hemispheric Sunspot Numbers. Moreover,
the scatter turns out to be rather small. In Table~\ref{correl}
the results of the regression analysis are summarized. The slope
derived from a linear least-squares fit to the data as well as the
cross-correlation coefficients are very close to 1. For the
standard error between the fitted and the original data we obtain
$\sim$3.7, given in units of the SIDC Sunspot Numbers. Thus, it
can be inferred that the derived daily hemispheric Sunspot Numbers
very well render the International ones.

On the basis of the daily $R_{n {\rm , KSO}}$ and $R_{s {\rm ,
KSO}}$ we derived also the monthly mean hemispheric Sunspot
Numbers for the period 1975--2000. In this regard, it has to be
stressed that the KSO data set does not steadily cover the overall
period; rather, 27\% of the daily values are missing. Thus, to
reconstruct those missing data we performed a linear interpolation
on the daily values, separately for the northern and southern
hemisphere, respectively. The middle panels in Fig.~\ref{Scatter}
show the scatter plots of the derived monthly mean northern
(southern) Sunspot Numbers against the corresponding International
northern (southern) Sunspot Numbers provided by the SIDC for the
overlapping period 1992--2000. In general, the derived monthly
data clearly follow the SIDC data, which is also reflected in the
high cross-correlation coefficients and the parameters of the
regression line (cf. Table~1). However, for the northern Sunspot
Numbers one outlier shows up (May 2000), in which the observed
daily data are obviously non-representative of the monthly mean.
This case can easily be explained by the exceptional low number of
observation days (5). It has to be noted that during the
considered period of 26~years only for three months was the data
coverage less than 11~days.
\begin{table}
\centering \caption{Summary of the regression analysis of the KSO
and SIDC hemispheric Sunspot Numbers 1992--2000. The analysis was
performed for the daily, the monthly mean and the smoothed monthly
mean northern and southern Sunspot Numbers. We list the
cross-correlation coefficients ({\it Corr.}), the parameters
obtained from the linear least-squares fit ({\it const., slope}),
and the standard error between the fitted and original data ({\it StE}). }
\begin{tabular}{lcccc} \hline\hline
         & Corr. & \multicolumn{2}{c}{Linear Fit}       & StE \\
         &              & const.            & slope            &  \\  \hline
daily N  & 0.991        & +0.009$\pm$0.103   & 0.993$\pm$0.003   & 3.662 \\
daily S  & 0.991        & +0.127$\pm$0.104   & 1.002$\pm$0.003   & 3.668 \\
monthly N  & 0.992      & +0.336$\pm$0.428   & 0.996$\pm$0.012   & 2.792 \\
monthly S  & 0.998      & $-0.026$$\pm$0.218 & 1.003$\pm$0.006   & 1.414 \\
sm.mon. N  & 0.999      & $+0.150$$\pm$0.088 & 0.975$\pm$0.003   & 0.508 \\
sm.mon. S  & 0.999      & $-0.014$$\pm$0.054 & 1.003$\pm$0.002   & 0.303 \\ \hline
\end{tabular}
\label{correl}
\end{table}

To circumvent the problem that possible outliers of the monthly
mean data also influence the preceding and succeeding months when
calculating the smoothed monthly mean hemispheric Sunspot Numbers,
we reversed the sequence of averaging and smoothing the data.
First we smoothed the daily (interpolated values included)
hemispheric Sunspot Numbers with a 365 days running average.
Subsequently we calculated the monthly means of this
annually-smoothed daily data. The bottom panels in
Fig.~\ref{Scatter} show the scatter plots of the derived smoothed
monthly Sunspot Numbers for the northern and southern hemisphere
versus the smoothed monthly mean hemispheric Sunspot Numbers from
the SIDC. Both panels clearly reveal that the derived smoothed
data closely match the SIDC data (see also the outcome of the
regression analysis summarized in Table~1). In particular, the
influence of the outlier of the northern monthly mean Sunspot
Numbers is almost eliminated. In Figure~\ref{Match/365} we plot
the time evolution of the derived smoothed monthly mean
hemispheric Sunspot Numbers (solid line) for the period
1992--2000, which closely resemble the corresponding data from the
SIDC (dashed line).
\begin{figure*}
 \resizebox{12cm}{!}{\includegraphics{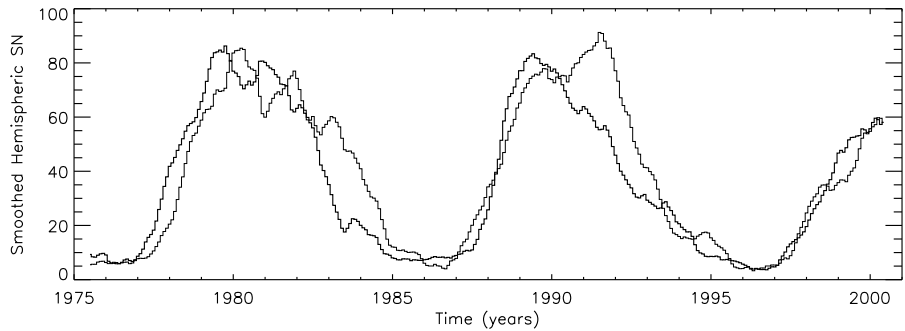}}
 \hfill
 \parbox[b]{55mm}{
 \caption{Smoothed monthly mean hemispheric Sunspot Numbers for the
    time span 1975--2000. The thick line indicates the northern, the thin line
    the southern Sunspot Numbers derived from the KSO data.}
 \label{result}
 }
\end{figure*}

\section{The catalogue}

The monthly mean and the smoothed monthly mean hemispheric Sunspot
Numbers for the period 1975--2000 are compiled into a
catalogue\footnote{More information concerning the catalogue as
well as access to the daily hemispheric sunspot data can be
requested by contacting: M.~Temmer.}. The catalogue is available
online at {\tt cdsarc.u-strasbg.fr} (130.79.128.5) or via {\tt
\small http://cdsweb.u-strasbg.fr/cgi-bin/qcat?J/A+A/390/707}. The
organization of the catalogue is described in the following
(for details see the {\tt ReadMe} file).
\begin{itemize}
\item{\bf 1st column:} Year and month (Months with less than
11 observation days are marked with an asterisk ($\ast$) to note
that the statistical significance of the derived monthly data is
low.)
\item{\bf 2nd column:} Monthly mean northern Sunspot Numbers
\item{\bf 3rd column:} Monthly mean southern Sunspot Numbers
\item{\bf 4th column:} Smoothed monthly mean northern Sunspot Numbers
(365 days running average followed by monthly mean calculations)
\item{\bf 5th column:} Smoothed monthly mean southern Sunspot Numbers
(365 days running average followed by monthly mean calculations)
\end{itemize}

\section{North-South asymmetry}

In Figure~\ref{result} the derived smoothed monthly mean northern
and southern Sunspot Numbers for the period 1975--2000 are shown.
From the figure it is obvious that the activity during the solar
cycle is not symmetric for both hemispheres. For instance, cycle
21 shows a clear phase shift between the northern and the southern
hemispheric activity. The existence of a N-S asymmetry has been
established and analyzed in several studies for a variety of solar
activity phenomena (e.g., flares, prominences, bursts, coronal
mass ejections, long-lived filaments, etc.). However, the physical
causes are still not satisfactorily interpreted. The analyses of
N-S asymmetries in the appearance of sunspots have mostly been
made on the basis of sunspot areas and sunspot group numbers
(Newton \& Milsom~\cite{Newton};
Waldmeier~\cite{Wald1},~\cite{Wald3}; White \&
Trotter~\cite{White}; Yallop \& Hohenkerk~\cite{Yallop}; Vizoso \&
Ballester~\cite{Vizoso}; Carbonell et al.~\cite{Carbonell}; Oliver
\& Ballester~\cite{Oliver}; Li et
al.~\cite{Li1},~\cite{Li2},~\cite{Li3}). Studies based on relative
sunspot numbers have been carried out by Swinson et al.
(\cite{Swinson}) who used data provided by Koyama (\cite{Koyama})
for the period 1947--1983.

\begin{figure}
\center{
  \resizebox{0.95\hsize}{!}{\includegraphics{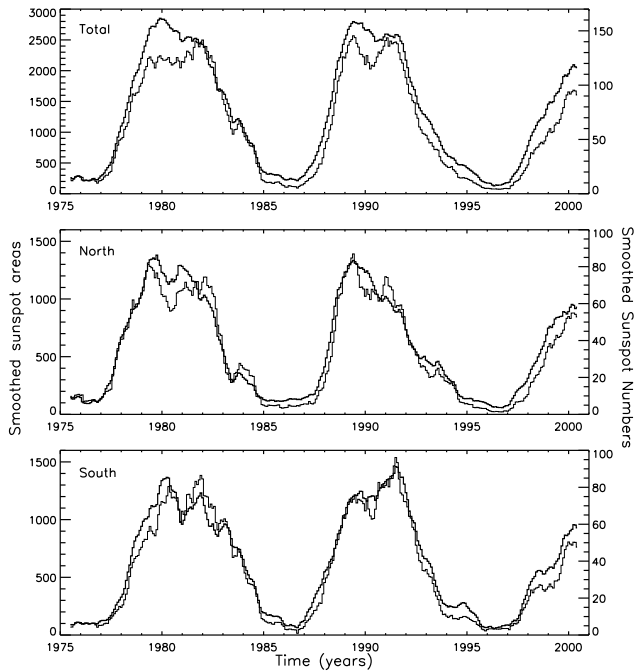}}
    }
     \caption{Smoothed monthly mean Sunspot Numbers and sunspot areas for the total Sun (top
     panel), the northern (middle panel) and the southern (bottom panel) hemisphere,
     respectively. Thin (thick) lines indicate hemispheric
     sunspot areas (Sunspot Numbers).}
\label{compare}
\end{figure}

In the following, we study the relation between Sunspot Numbers
and areas, the significance of the \mbox{N-S} asymmetry excesses
as a function of the solar cycle, and dominant rotational periods
for the northern and southern hemisphere.

\subsection{Sunspot Numbers -- Sunspot areas}

In Figure~\ref{compare}, we plot the smoothed monthly mean sunspot
areas (data are taken from the Royal Greenwich Observatory)
together with the derived Sunspot Numbers, for the whole Sun as
well as separately for the northern and southern hemisphere. The
figure clearly reveals that considering the activity of the
northern and southern hemisphere separately (areas as well as
Sunspot Numbers), different information is provided than when
considering the whole disk. For instance, in the hemispheric
indices, the activity gaps during the maximum phase, the so-called
Gnevyshev gaps (Gnevyshev~\cite{Gnevyshev}), are clearly visible,
whereas they are often smeared out when considering the total
activity. Furthermore, it can be clearly seen that the northern
and southern hemisphere do not reach their maximum simultaneously
but there may be a shift of up to several years (see in particular
cycle~22). Thus, the time as well as the height of the maximum and
the Gnevyshev gap in the total activity can be understood as a
superposition of both hemispheres, which provide the primary
physical information on solar activity.

For the cross-correlation coefficients between the monthly mean
sunspot areas and sunspot numbers, we obtain 0.90, 0.91 and 0.94
for the northern, the southern and the total component,
respectively, indicating a good correspondence between both
activity indices. However, the relationship is far from being
one-to-one. Especially during the maximum phase, significant
differences between sunspot numbers and areas appear (see
Fig.~\ref{compare}). In principle, sunspot areas are a more direct
physical parameter, being closely related to the magnetic field.
However, the reliable measurement of sunspot areas is not an easy
task, and the results derived by different techniques and
different observatories may differ by an order of magnitude
(Pettauer \& Brandt,~\cite{Pettauer}). This poses in particular
problems for mid- and long-term investigations of solar activity.
Furthermore, from an ongoing study we obtained that, for instance,
the hemispheric occurrence of H$\alpha$ flares is more closely
related to the Sunspot Numbers than to the sunspot areas, which
emphasizes the high physical relevance of Sunspot Numbers (Temmer
et al., in \mbox{preparation}).

\subsection{N-S Asymmetry: Excess}

\begin{figure}
\center{
  \resizebox{0.95\hsize}{!}{\includegraphics{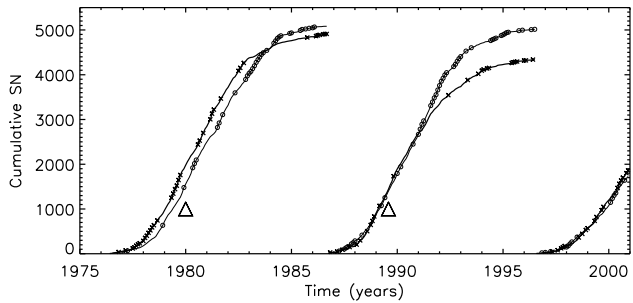}}
       }
     \caption{The cumulative monthly mean Sunspot Numbers for the northern (thick line) and
     southern (thin line) hemisphere, respectively, are presented
     separately for solar cycle 21, 22
     and the rising phase of solar cycle 23.
     Crosses (circles) indicate an excess of the northern
     (southern) hemisphere for the respective month on a 95\%
     significance level. Triangles indicate the maximum of the
     solar cycle.}
\label{cum}
\end{figure}

Figure~\ref{cum} shows the cumulative monthly mean northern and
southern Sunspot Numbers, separately plotted for solar cycles 21,
22 and the rising phase of the current cycle 23. In order to
assess the significance of the activity excess of the northern or
southern hemisphere, respectively, we applied the paired Student's
t-test. The test statistics $\hat{t}$ is defined by
\begin{equation}\label{t-test}
\hat{t} = \frac{\bar{D}}{s_{\bar{D}}} = \frac{(\Sigma
D_{i})/n}{\sqrt{\frac{\Sigma D_{i}^{2} - (\Sigma D_{i})^{2}/n}{n(n
- 1)}}}  \, ,
\end{equation}
where $D_{i}$ is the difference of paired values (here, daily
$R_{n}$,$R_{s}$), $\bar{D}$ the mean of a number of $n$
differences and $s_{\bar{D}}$ the respective standard deviation
with $n-1$ degrees of freedom. Since we want to test the
significance of the monthly value, $n$ is given by the number of
days of the considered month. The calculated test value,
$\hat{t}$, based on the degrees of freedom is compared to the
corresponding $\hat{t}_{n-1, \alpha}$ given in statistical tables
on a preselected error probability $\alpha$. We have chosen
$\alpha=0.05$, i.e. if $\hat{t}>\hat{t}_{n-1, \alpha}$, the
difference between the paired values is statistically significant
at a 95\% level. Thus for each month, the paired Student's t-test
is utilized to determine the significance of the difference
between the northern and southern Sunspot Numbers.
The results of this test are indicated in the graph of the
cumulative Sunspot Numbers (Fig.~\ref{cum}). The calculated
$\hat{t}$ values with 95\% significance are overplotted at each
specific month signed as crosses (circles) for the northern
(southern) hemisphere to represent that the excess of the flagged
hemisphere is highly significant. This representation shows that
during solar cycle 21 the excess of the southern hemisphere has
significant excesses predominantly at the end of the cycle,
whereas the northern hemisphere is more active at the beginning
and the maximum phase. For solar cycle 22, both hemispheres show a
similar amount of activity excesses during the ascending phase,
and almost no predominance of one hemisphere over the other during
the maximum phase. The distinct excess of the southern activity is
exclusively built up during the declining phase. The current solar
cycle 23 is only covered by its rising phase with a slight excess
of northern activity. Similar results for solar cycle 21 based on
cumulative counts of soft X-ray flares are reported by
Garcia~(\cite{Garcia}), and for solar cycle 21 and 22 on the basis
of H$\alpha$ flares by Temmer et al.~(\cite{Temmer}).

In Table~\ref{test} the outcome of the Student's t-test is
summarized, listing the percentages of significant months in
relation to the total number of months during the specific solar
cycles. In each cycle, about 60\% of all months reveal a highly
significant N-S asymmetry. For solar cycle 21, the percentage of
months with significant activity excess is slightly higher for the
northern than for the southern hemisphere. For solar cycle 22, the
southern hemisphere covers almost twice as much significant months
than the northern (cf. Table~\ref{test}). Regarding the total
activity during the cycle, we obtain a slight excess of the
southern hemisphere over the northern hemisphere with about 50.9\%
during solar cycle 21 and a distinct excess of 53.6\% during solar
cycle 22 (cf. Fig.~\ref{cum}).

Swinson et al. (\cite{Swinson}) who analyzed Sunspot Numbers for
the period 1947 until 1983 almost completely including solar
cycles 19, 20 and 21, show good agreement with our results,
concluding that the northern hemisphere peaks in its activity
excess about two years after solar minimum. These authors report
that this peak is greater during even cycles which points out a
relation to the 22 year solar magnetic cycle. Balthasar \&
Sch\"ussler (\cite{Balthasar1},~\cite{Balthasar2}) interpreted the
distribution of daily relative sunspot numbers as a kind of solar
memory with preferred hemispheres that alternate with the 22 year
magnetic cycle. Contrary to that, Newton \& Milsom (\cite{Newton})
and White \& Trotter (\cite{White}), by analyzing sunspot areas,
did not find a systematic change in activity between both
hemispheres, i.e. no evidence for a dependence on the 22 year
solar magnetic cycle.

\begin{table}
\centering \caption{The percentage of months ($T$) with 95\%
significant N-S asymmetry with respect to the total number of
months for solar cycles 21, 22 and the rising phase of solar cycle
23 is given. Additionally, the significant months are subdivided
into the northern ($N$) and southern ($S$) hemisphere.}
\begin{tabular}{ccccc} \hline\hline
 Cycle No. & $T(\%)$ & $N(\%)$ & $S(\%)$\\ \hline
 21  & 59.3  & 30.9 & 28.4  \\
 22  & 63.2  & 22.2 & 41.0  \\
 23  & 60.3  & 35.8 & 24.5  \\ \hline
\end{tabular}
\label{test}
\end{table}

\subsection{Rotational periods}

Various periodicities have been detected in solar activity time
series. The most prominent are the 11 year solar cycle and the
27~day Bartels rotation (Bartels \cite{Bartels}). In the present
study, we are interested in periods related to the solar rotation
and in their behavior with regard to the N-S
occurrence.\footnote{Note: All periods quoted in this paper are
synodic.} For this purpose, we analyze power spectra and
autocorrelation functions from daily Sunspot Numbers of the
northern hemisphere, the southern hemisphere and the total solar
disk.

For the power spectrum analysis, we have adopted the Lomb-Scargle
periodogram technique (Lomb,~\cite{Lomb}; Scargle,~\cite{Scargle})
modified by Horne \& Baliunas (\cite{Horne}). By this method, the
power spectral density (PSD) is calculated normalized by the total
variance of the data. This periodogram technique is particularly
useful in order to assess the statistical confidence of a
frequency identified in the periodogram by computing the false
alarm probability ($FAP$). In our case, the relevant time series
are prepared from the daily Sunspot Numbers, which are not
independent of each other but correlated with a typical
correlation time of about 7 days (Oliver \&
Ballester~\cite{Oliver2}). Thus, the statistical significance of a
peak of height~$z$ in the periodogram has to be tested for the
case that the data are statistically correlated. Therefore, the
PSD has to be normalized by a correction factor $k$, which should
be determined empirically (described below). Then the $FAP$ can be
derived by the following equation
\begin{equation}\label{fap}
FAP = 1 - [1 - \exp(-z_{m})]^{N} \, ,
\end{equation}
where $N$ denotes the number of independent frequencies in the
time series, $z_{m}=z/k$ is the derived normalized power, $z$ is
the Scargle power and $k$ the normalization factor due to event
correlation (Scargle~\cite{Scargle}; Horne \&
Baliunas~\cite{Horne}; Bai \& Cliver~\cite{Bai2};
Bai~\cite{Bai3}).

The number of independent frequencies is given by the spectral
window investigated and the value of the independent Fourier
spacing, $\Delta f_{\rm ifs}=1/\tau$, where $\tau$ is the time
span of the data (Scargle~\cite{Scargle}). Here, we are interested
in periods related to the Sun's rotation, for which we have chosen
the spectral window [386,463] nHz, i.e. [25,30] days. Considering
a time span from January 1975 to December 2000 we have $\tau=9497$
days, hence $\Delta f_{\rm ifs}=1.22$ nHz. However, de Jager
(\cite{Jager}) has shown by Monte-Carlo simulations that the
Fourier powers taken at intervals of one-third of the independent
Fourier spacing are still statistically independent, i.e
$\frac{\Delta f_{\rm ifs}}{3}=0.41$ nHz. Thus we accepted the
number $N=190$ as the number of independent frequencies in the
chosen spectral window.

\begin{figure}
\center{
 \resizebox{0.88\hsize}{!}{\includegraphics{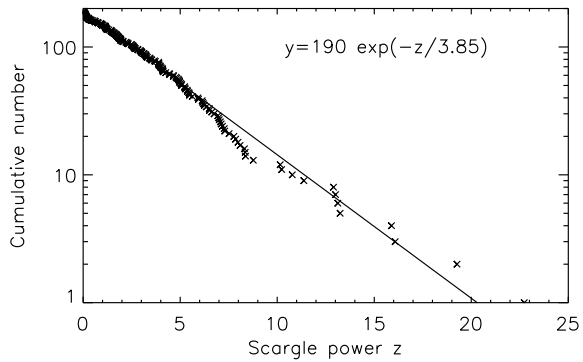}}
     }
     \caption{The cumulative distribution function of the power
     values derived from the Lomb-Scargle periodogram for the total solar disk.
     The y-axis shows the cumulative number of frequencies whose power
     exceeds a height $z$ and the x-axis the values of power, $z$.
     The solid line shows the best fit to the distribution for $z<7$.}
\label{scargle}
\end{figure}

We briefly describe the method used to calculate the normalization
factor $k$. The normalization factor $k$ is derived from the
cumulative distribution function of the Scargle power values $z$
for the 190 independent frequencies (shown for the Sunspot Numbers
of the total solar disk in Fig.~\ref{scargle}). The distribution
can be well fitted to the equation $y=190\exp(-z/k)$ for power
values $z<7$, which gives the normalization factor $k$ (indicated
in Fig.~\ref{scargle} as solid line). For the time series of the
total solar disk we obtain $k=3.85$, thus the power spectrum is
normalized once more by dividing the Scargle power by 3.85. For
the northern hemisphere we obtain $k=6.63$ and for the southern
hemisphere $k=5.18$. Further detailed descriptions of this
procedure can be found in Bai \& Cliver (\cite{Bai2}), Oliver \&
Ballester~(\cite{Oliver2}) and Zi\c{e}ba et al. (\cite{Zieba}).

\begin{figure}
\center{
 \resizebox{0.87\hsize}{!}{\includegraphics{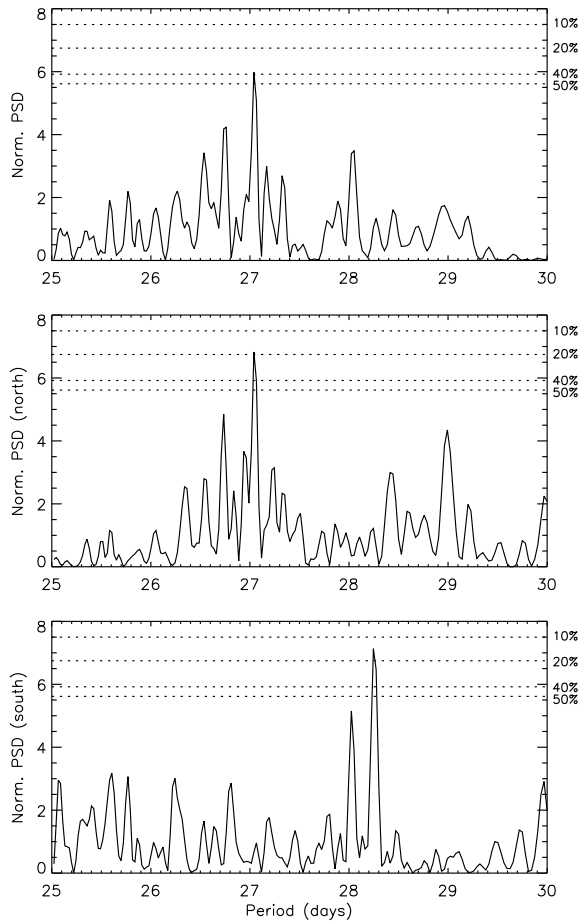}}
      }
     \caption{Periodograms derived from the daily Sunspot Numbers
     of the total solar disk (top panel), the northern hemisphere
     (middle panel) and the southern hemisphere (bottom
     panel), from the time span 1975--2000. The dashed lines
     indicate various $FAP$ levels.}
\label{power}
\end{figure}

The normalized PSD for the total and the hemispheric Sunspot
Numbers are represented in Fig.~\ref{power}. The $FAP$ levels
calculated by Eq.~\ref{fap} are indicated as dashed lines for the
probabilities of 50\%, 40\%, 20\% and 10\%. As can be seen, the
total Sunspot Numbers show only one peak above the 50\%
significance level, namely at 27.0 days with a $FAP$ value of
38\%. The northern Sunspot Numbers show also one peak above the
50\% significance level at 27.0 days with a $FAP$ value of 19\%.
For the southern Sunspot Numbers we get one peak at 28.2 days
which lies above the 50\% significance level with a $FAP$ value of
12\%. Thus, for the northern hemisphere the power is mainly
concentrated at $\sim$27~days, whereas for the southern hemisphere
it is mainly concentrated at $\sim$28~days. These rotational
properties manifest a strong asymmetry with respect to the solar
equator. It is worth noting that the spectral power of the peaks
is lower for the total Sunspot Numbers than for the hemispheric
ones. This phenomenon probably arises from the fact that in the
case of the total Sunspot Numbers, the components of both
hemispheres, which are obviously not in phase, overlap and result
in a lower PSD than for the hemispheric Sunspot Numbers.
Since there is no enhanced PSD at 27~days for the time series of
the southern Sunspot Numbers, the signal of the 27 days Bartels
rotation found for the total Sunspot Numbers is a consequence of
the northern component, in which the 27 days peak is very strong.
On the other hand, the enhanced PSD at $\sim$28 days, found for
the southern Sunspot Numbers, is not powerful enough to be
reflected as a significant period in the PSD of the total Sunspot
Numbers (see Fig.~\ref{power}, top panel).

\begin{figure}
\center{
 \resizebox{0.89\hsize}{!}{\includegraphics{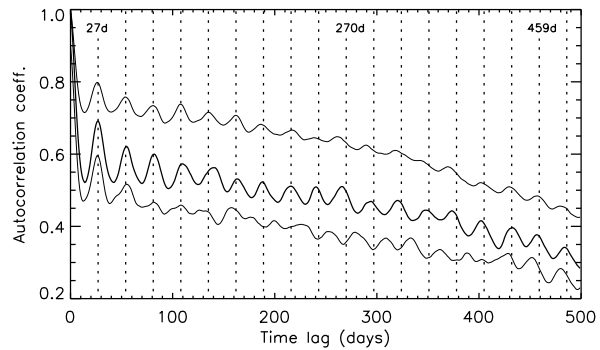}}
      }
     \caption{Autocorrelation function of the daily Sunspot Numbers
     for the period 1975--2000, plotted up to a time lag of 500 days.
     The top line indicates the autocorrelation function for the
     total disk, the thick line for the northern and
     the thin line for the southern hemisphere. The
     27~days Bartels rotation up to 15 periods is overplotted by dotted
     lines. (For better representation, the autocorrelation
     function of the southern Sunspot Numbers is shifted
     downwards by a factor~$0.1$.)}
\label{auto}
\end{figure}

In Fig.~\ref{auto}, we show the autocorrelation function derived
from the northern, the southern and the total daily Sunspot
Numbers, up to a time lag of 500~days. Distinct differences appear
for the behavior of the northern and the southern hemisphere. The
northern Sunspot Numbers reveal a stable periodicity of about 27
days, which can be followed for up to 15~periods. Contrary to
that, the signal from the southern hemisphere is strongly
attenuated after 3~periods, and shows a distinctly less regular
periodicity. The autocorrelation function of the total Sunspot
Numbers reveals an intermediate behavior, resulting from the
superposition of the northern and southern \mbox{components}.

\section{Discussion}

Bogart (\cite{Bogart}) investigated autcorrelation functions of
daily Sunspot Numbers for the period 1850--1977, obtaining a
distinct period at 27 days. Since the persistence of this
periodicity lasted for about 8--12 solar rotations, which is
distinctly longer than the lifetimes of sunspots, Bogart
(\cite{Bogart}) concluded that this outcome provides evidence for
the existence of ``active longitudes" on the Sun. The idea that
solar activity phenomena (sunspots, flares, etc.) are not
uniformly distributed in longitude but preferentially occur at
certain longitude intervals was already suspected by Carrington
(\cite{Carrington}), and has been systematically studied since the
1960s (e.g., Trotter \& Billings~\cite{Trotter},
Warwick~\cite{Warwick}, Bumba \& Howard~\cite{Bumba}). In later
studies it became clear that these ``active longitudes" usually do
not extend over both hemispheres, and the terms ``active zones"
and ``complexes of activity" are often used instead.

In the present analysis, studying the period 1975--2000, we
obtained a similar outcome for the autocorrelation function of the
daily Sunspot Numbers as Bogart (\cite{Bogart}). On the other
hand, the supplementary analysis of the hemispheric Sunspot
Numbers reveals that the persistent 27 day period occurs only for
the northern hemisphere. The autocorrelation function of the
southern Sunspot Numbers is strongly attenuated after 3~periods,
which matches with the lifetimes of long-lived sunspot groups. The
autocorrelation function derived from the northern Sunspot Numbers
shows a stable periodicity up to 15~periods. This outcome provides
strong evidence that during the considered period a preferred zone
of \mbox{activity} was present at the northern hemisphere rigidly
rotating with 27~days, whereas the behavior of the southern
hemisphere is mainly dominated by individual long-lived sunspot
groups, which are not (or only weakly) grouped at preferred
longitude intervals.

Bai (\cite{Bai1}) studied the occurrence of major solar flares
during 1980--1985, and found evidence for a very prominent active
zone in the northern hemisphere, in which almost half of the
flares of the northern hemisphere took place (see his Fig.~3), and
a second (but weaker) one separated by~$\sim$180$^\circ$. In the
periodograms, we find a highly significant peak at
$\sim$13.3~days, which gives indication for a second active
zone~$180^\circ$ apart. Such a peak was noticed too, e.g., by
Bogart (\cite{Bogart}). Bai (\cite{Bai1}) reports also evidence
for two active zones in the southern hemisphere, but with a
distinctly less pronounced activity than the major northern active
zone. In the present study we do not find evidence for an active
zone in the southern hemisphere. A possible explanation is that
the hemispheric Sunspot Numbers, which represent a quantity
averaged over a whole hemisphere, are less sensitive to the
detection of active zones than the binning of solar activity
phenomena in longitude intervals, as done by Bai (\cite{Bai1}).

The 27 day Bartels rotation is a very prominent period with
respect to the occurrence of geomagnetic disturbances, and it is
supposed to be relevant for the large scale solar magnetic fields
(e.g. Balthasar \& Sch\"ussler~\cite{Balthasar2}). From the
distribution of daily Sunspot Numbers and sunspot groups in the
Bartels rotation of 27~days, Balthasar \& Sch\"ussler
(\cite{Balthasar1},~\cite{Balthasar2}) inferred evidence for
preferred Bartels longitudes of activity, which may even cover a
whole hemisphere. These preferred hemispheres were found to
alternate with the 22~year magnetic cycle. The present data set
(1975--2000) is too short to draw any conclusion with respect to
the 22~year magnetic cycle. However, it is worth noting that
during the considered period the total activity is higher in the
southern hemisphere (see Fig.~\ref{cum}), but the northern
hemisphere shows a distinctly more systematic behavior in the
rotational recurrence of activity (see Fig.~\ref{auto}), very
probably dominated by one or two preferred active zones. For solar
cycle 21, a cyclic behavior of the N-S asymmetry with a phase
shift between both hemispheres can be inferred from the cumulative
Sunspot Numbers (Fig.~\ref{cum}) as well as directly from
Fig.~\ref{result}, whereas no cyclic behavior is indicated for
solar cycle 22. Nevertheless, a cyclic behavior in the \mbox{N-S}
asymmetry of solar activity for the last 8~solar cycles was
recently reported by Vernova et al. (\cite{Vernova}), using the
so-called ``vectorial sunspot area", which more strongly
emphasizes the systematic, longitudinally asymmetric sunspot
activity compared to the stochastic, longitudinally evenly
distributed component than the normal sunspot areas and Sunspot
Numbers.

The study of N-S asymmetries of solar activity and the analysis of
the rotational behavior separately for the northern and southern
hemisphere is particularly relevant, as it is related to the solar
dynamo and the generation of magnetic fields. Antonucci et al.
(\cite{Antonucci}) investigated the rotation of photospheric
magnetic fields during solar cycle 21, and obtained a dominant
period of 26.9 days for the northern and 28.1 days for the
southern hemisphere. The spectral power was concentrated in a few
well-defined regions with a rather wide extent in latitude. On
average, these regions were found to lie near the sunspot
differential curve, but the latitudinal extent was not consistent
with the standard differential curve. From these findings,
Antonucci et al. (\cite{Antonucci}) concluded that the emergence
of photospheric magnetic field tracers is organized in a
large-scale pattern with different rotation periods in both
hemispheres.

The rotation rates for the magnetic fields of the northern and
southern hemisphere reported by Antonucci et al.
(\cite{Antonucci}) coincide well with the present study of
hemispheric Sunspot Numbers, in which we found an enhanced PSD at
$\sim$27~days with a rigid rotation for the northern hemisphere,
and $\sim$28~days for the southern hemisphere. Thus, a strong
correlation of the underlying large-scale magnetic fields and the
rotation of sunspots is suggested. It has to be noted that
Nesme-Ribes et al. (\cite{Nesme}), who analyzed sunspots observed
on spectroheliograms during solar cycle 21, came to an opposite
conclusion, since the 28~days period in the southern hemisphere
could not be reproduced. In this regard, it has to be stressed
that the rotational behavior of the solar surface is sensitive to
various factors, such as, for instance the used tracers (e.g., old
versus young sunspots) as well as the phase of the solar cycle
(e.g., Balthasar et al.~\cite{Balthasar3}, Nesme-Ribes et al.
\cite{Nesme}, Pulkkinen \& Tuominen \cite{Pulkkinen}, and
references therein).

The existence of significant N-S asymmetries in the occurrence of
solar activity and in the rotational behavior provides strong
evidence that the magnetic field systems originating in the two
hemispheres are only weakly coupled (see also Antonucci et al.
\cite{Antonucci}). Another feature that gives indications for a
weak coupling of the hemispheric activities and the related
magnetic field evolution within the course of a solar cycle is the
double-peaked cycle maximum known since Gnevyshev
(\cite{Gnevyshev}), which is supposed to be related to the solar
magnetic field reversal (e.g., Feminella \&
Storini~\cite{Feminella}, Bazilevskaya et al.~\cite{Bazil}, and
references therein). From Fig.~\ref{compare} it can be seen that
the Gnevyshev gap is more pronounced by considering the northern
and southern Sunspot Numbers (or sunspot areas) separately and it
does not necessarily take place at the same time and with the same
strength in both hemispheres.

In principle, studies of the N-S asymmetry of solar activity
provide constraints on solar dynamo theories. In addition to the
11~year sunspot cycle, the 22~year magnetic cycle, the presence of
grand sunspot minima, the butterfly diagram and phase-amplitude
correlations, a reliable dynamo theory should be able to explain
the weak coupling between the hemispheres and the existence of
significant N-S asymmetries in the occurrence of solar activity as
well as in the rotational behavior. Ossendrijver et al.
(\cite{Ossen}) have shown that a mean field dynamo model with
stochastic fluctuations of~$\alpha$ can reproduce observed N-S
asymmetries in solar activity. A probable candidate for the origin
of these stochastic fluctuations are giant convective cells that
extend sufficiently deep into the convection zone (Ossendrijver et
al. \cite{Ossen}). The existence of giant cells was first
suggested by Bumba \& Howard (\cite{Bumba}), but their subsequent
history was quite controversial (see Hathaway et
al.~\cite{Hathaway}). Recently, Beck et al. (\cite{Beck}) reported
evidence for long-lived giant velocity cells on the solar surface
that extend $40-50^{\circ}$ in longitude but less than
$10^{\circ}$ in latitude. The magnetic fields of solar activity
are generated at the bottom of the convection zone, and long-lived
cells connected to this layer could explain the persistence of
solar activity at the same location (Beck et al.~\cite{Beck}).
Thus, giant convective cells possibly provide the physical link
between the solar interior and active zones at the solar surface,
whose existence has been established for various solar activity
phenomena and for various solar cycles. To account for the
observational effects, these active zones must not be
significantly disrupted by differential rotation up to several
years. On the other hand, as suggested by Balthasar \& Sch\"ussler
(\cite{Balthasar1}), the rigid 27 days rotation does not
necessarily represent any material motion but may result from the
phase velocity of a dynamo-produced large-scale solar magnetic
structure.

\section{Conclusions}

Hemispheric Sunspot Numbers, in addition to the historical Sunspot
Numbers that describe the activity of the whole Sun, provide
important information on solar activity. Distinct asymmetries
between the northern and the southern hemisphere appear concerning
various aspects of solar activity: the location and height of the
cycle maximum and the Gnevyshev gap, the alternating predominance
of one hemisphere over the other, and the different rotational
behavior of activity tracers in both hemispheres. These findings
suggest that the magnetic field systems originating in the two
hemispheres and their evolution in the course of the solar cycle
are only weakly coupled. Finally, we stress that the various
phenomena of N-S asymmetry in solar activity provide valuable
constraints for solar dynamo theories.

\acknowledgements The authors thank the former and present staff
of the KSO for performing and making available the sunspot
drawings, and also for helpful comments. We are grateful for the
thoughtful comments of the referee, H. Balthasar. M.T., A.V. and
A.H. gratefully acknowledge the Austrian {\em Fonds zur
F\"orderung der wissenschaftlichen Forschung} (FWF grants
P13653-PHY and P15344-PHY) for supporting this project.

\end{document}